\documentclass[%
reprint,
superscriptaddress,
 amsmath,amssymb,
prb,
]{revtex4-2}

\usepackage[dvipdfmx]{graphicx}% Include figure files
\usepackage{dcolumn}% Align table columns on decimal point
\usepackage{bm}% bold math
\usepackage{braket}
\usepackage[export]{adjustbox} 

\DeclareMathOperator{\Tr}{Tr}
\usepackage{multirow}
\usepackage{amsmath}
\usepackage{comment}
\usepackage{mathrsfs}

\usepackage[unicode, colorlinks, citecolor=blue, linkcolor=red, urlcolor=blue]{hyperref} %, breaklinks=true

\begin{document}

%\preprint{APS/123-QED}
%\title{Manuscript Title:\\with Forced Linebreak}% Force line breaks with \\

%\affiliation{%

%\affiliation{
%\affiliation{
%\affiliation{%

\title{{Magnetoresistance in chiral systems driven by inter-band spin-orbit coupling}} 

\author{Misa Nozaki}
 \email{6d697361@gmail.com}
\affiliation{Institute for Quantum Life Science, National Institutes for Quantum Science and Technology (QST), Chiba 263-8555, Japan}

\author{Takatoshi Fujita}
 \email{fujita.takatoshi@qst.go.jp}
\affiliation{Institute for Quantum Life Science, National Institutes for Quantum Science and Technology (QST), Chiba 263-8555, Japan} 


\date{\today}

\begin{abstract}
Chiral-induced spin selectivity (CISS), in which electrons transmitted through nonmagnetic chiral materials exhibit strong spin-dependent transport, has attracted growing interest for spintronic applications.
However, a quantitative understanding of CISS remains elusive, partly because most previous studies rely on single-band models. 
In this work, we theoretically investigate multi-band effects on magnetoresistance (MR)-CISS, which is typically observed in experiments using magnetic conductive atomic force microscopy. 
To evaluate the spin polarization in MR-CISS, we simulate the nonequilibrium steady-state current using the Gorini-Kossakowski-Sudarshan-Lindblad master equation. 
We find that spin polarization exceeding 25\% can be achieved for realistic inter-band spin-orbit coupling strengths in the presence of on-site Coulomb interactions. 
These findings highlight the crucial role of inter-band spin-orbit coupling in the mechanism of CISS.
\end{abstract}

\maketitle

\section{Introduction} 
Chiral symmetry in matter has recently been recognized as a key origin of spin-dependent responses.
This understanding can be traced back to a photoelectron spectroscopy study on L- or D-stearoyl lysine in 1999, which provided the first evidence that chiral materials can exhibit spin filtering \cite{Ray1999}. 
Such spin-dependent phenomena are now referred to as chiral-induced spin selectivity (CISS) and have been observed in a wide range of non-magnetic chiral systems, including organic molecules, polymers, proteins, and metal oxides, as well as in diverse physical processes such as photoemission \cite{Gohler2011_Science, Naaman2015_ACIE, Kettner2018, CuO2019, CuO2022}, photoinduced charge separation \cite{Eckvahl2023_science, Eckvahl2024_JACS, Latawiec2025_PNAS}, electron transport \cite{JCP_MR_Kiran_2017, NatComm_MR_Suda_2019, JPCC_MR_Mishra_2020, AdvMater_Kulkarni_2020, JPCC_Das_2022, JACS_Rodriguez_2022, JACS_Zhang_2023, NatNanotech_MR_Saito_2025}, and superconductivity \cite{PRX_sato_2025}. 
CISS is currently expected to have potential applications in areas such as spintronic devices \cite{Mishra2025_JMCC, NatRevPhys_SHYang_2021}, electrochemical energy systems (e.g., metal-air batteries) \cite{Photoemission_review_2026}, and enantiomer separation systems \cite{Koyel2018_science}.

Despite extensive experimental evidence for CISS, its underlying mechanism remains unresolved.
Early theoretical studies proposed model Hamiltonians with helical symmetry and spin-orbit coupling (SOC), which can give rise to spin polarization \cite{PRL_Guo_2012, PNAS_Guo_2014}. 
However, the SOC strength in typical organic systems is generally considered insufficient to account for the observed CISS effects. Consequently, recent studies have focused on identifying additional mechanisms that enable a quantitative description of CISS \cite{Hanggai_ARPC_2026}.

To elucidate CISS mechanisms, spin polarization of electrons in two-terminal systems has been extensively studied both experimentally and theoretically. Experiments are typically performed using magnetic conductive atomic force microscopy (mc-AFM). Many experiments have shown that the magnetization direction of the AFM tip affects the magnitude of the current flowing through a chiral material sandwiched between a magnetic tip and a nonmagnetic metal substrate. This phenomenon is sometimes referred to as magnetoresistance (MR) CISS. MR-CISS has been reported in systems such as DNA, alkenes, helicenes, and organic molecular aggregates \cite{JCP_MR_Kiran_2017,NatComm_MR_Suda_2019,JPCC_MR_Mishra_2020,AdvMater_Kulkarni_2020,JPCC_Das_2022,JACS_Rodriguez_2022,JACS_Zhang_2023,NatNanotech_MR_Saito_2025}.

So far, several theoretical studies have revealed the importance of electron correlation in reproducing experimentally observed MR-CISS properties \cite{JPhysChemLett_Fransson_2019, JPCC_Huisman_2022, Meng2024_PRB}. 
Fransson proposed a model Hamiltonian incorporating on-site Coulomb interaction as well as chiral symmetry and SOC, which can give rise to significant spin polarization \cite{JPhysChemLett_Fransson_2019}. Using the same model, Huisman \textit{et al.} qualitatively reproduce experimentally observed bias voltage dependence of spin polarization \cite{JPCC_Huisman_2022}. Furthermore, Fransson and co-workers showed that the temperature dependence observed in experiments can be explained based on electron-phonon interactions and on-site Coulomb interactions \cite{PhysRevB_Fransson_2020, NanoLett_Fransson_2021, JPCC_Das_2022}.

 However, further extensions of these models are required to quantitatively understand MR-CISS. In the previous theoretical studies \cite{JPhysChemLett_Fransson_2019, JPCC_Huisman_2022, Meng2024_PRB}, helically oriented atoms or molecules are considered using Hubbard-type models that incorporate SOC between nearest-neighbor or next-nearest-neighbor sites. Such models, however, are not suitable for describing MR-CISS in helical molecular aggregates \cite{AdvMater_Kulkarni_2020, NatNanotech_MR_Saito_2025} because SOC between spatially distant orbitals is small or nearly negligible. Utsumi \textit{et al.} analyzed the spin-filtering effect using a multi-band model (two orbitals per site) including intra-site SOC, which is more appropriate for helical molecular aggregates, but their study does not address MR-CISS \cite{PhysRevB_Utsumi_2020, IsrJChem_Utsumi_2022, EurPhysJSpecTop_Kato_2025}.

In this work, we investigate MR-CISS using a two-band, two-site extended Hubbard model with intra-site SOC, focusing on the role of multi-band effects.
Using the Gorini-Kossakowski-Sudarshan-Lindblad (GKSL) master equation \cite{GKS, Lindblad}, we simulate the current over a wide range of on-site Coulomb interaction strengths. 
Our results reveal spin polarization exceeding 25\% and show that intra-site SOC plays a crucial role in generating MR-CISS in the presence of on-site Coulomb interactions.

\begin{figure}[htbp]
\label{fig:1}
\begin{tabular}{l}
(a)\\
\includegraphics[bb=0 0 429 297,width=0.65\linewidth,valign=c]{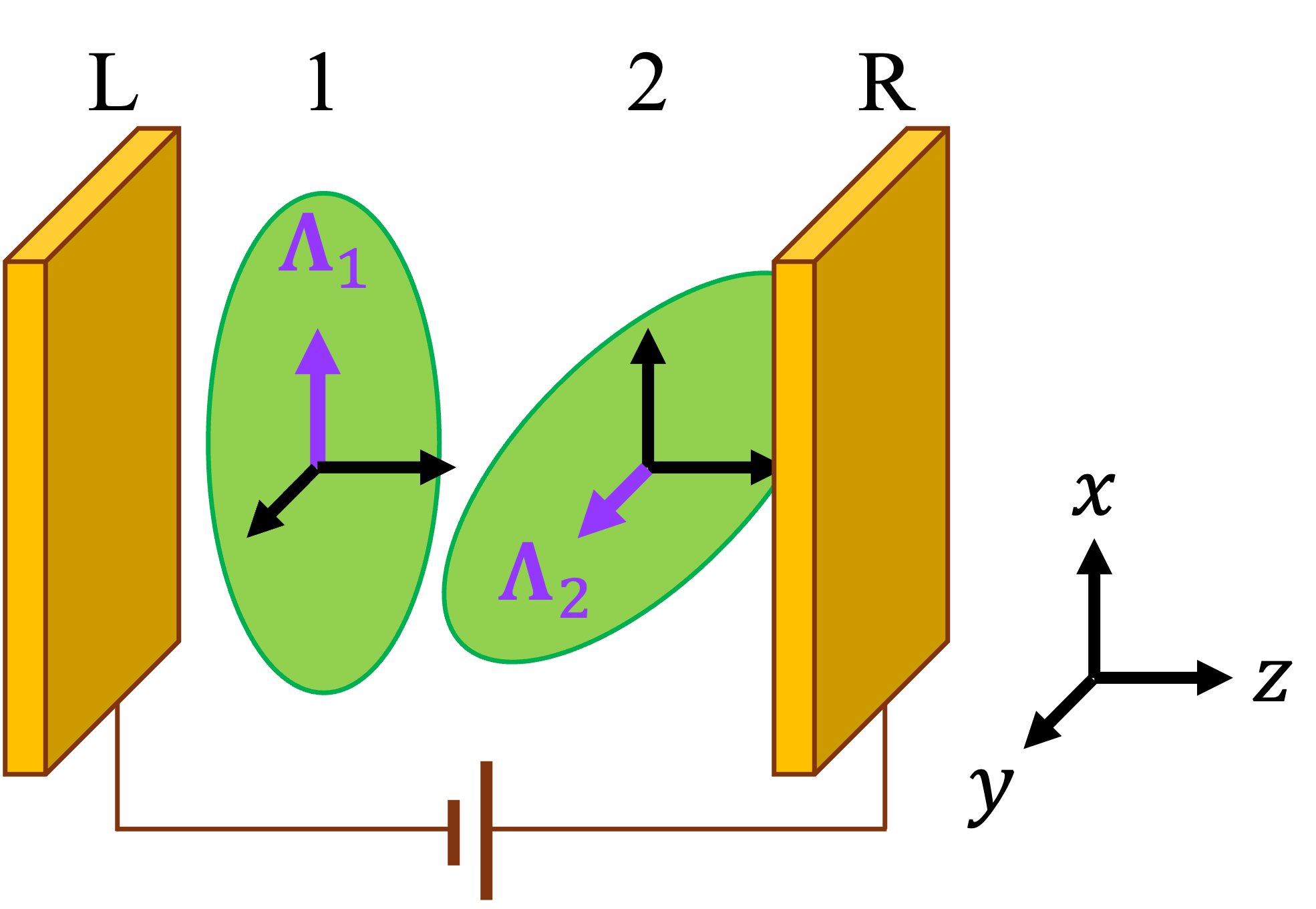} \\
(b)\\
\includegraphics[bb=0 0 392 420,width=0.62\linewidth, valign=c]{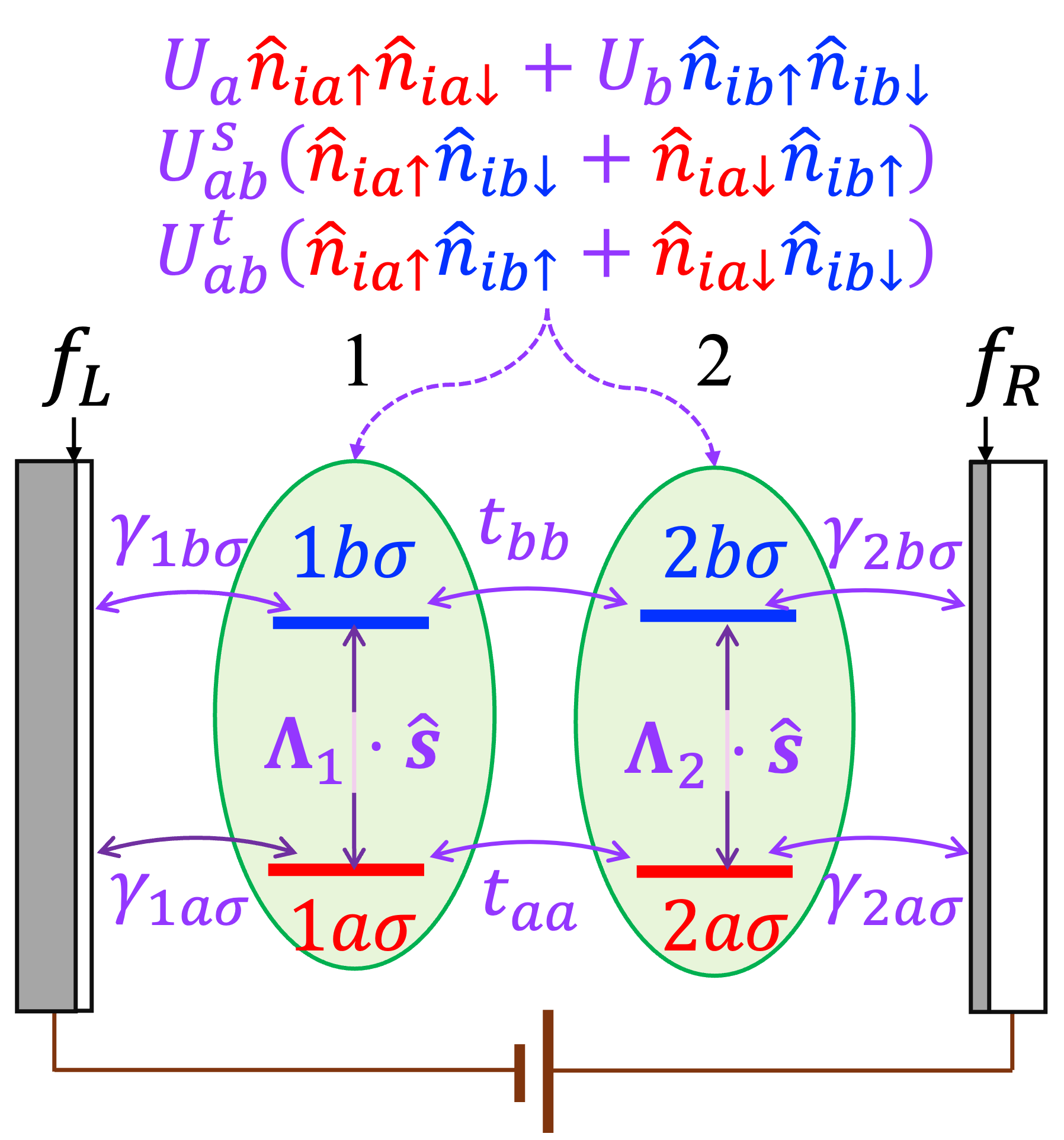}\\
\end{tabular}
\caption{
(a) Schematic illustration of the model. 
The two yellow rectangles (labeled $L$ and $R$) represent the left and right electrodes. 
The two green circles labeled $1$ and $2$ represent identical molecules. 
%The purple arrows labeled $\bm{\Lambda}_i=\bra{ia}\hat{\bm \Lambda}\ket{ib}$ represent the matrix elements of orbital parts of the one-electron SOC operator. 
The purple arrows labeled $\bm{\Lambda}_i=\bra{ia}\hat{\bm \Lambda}\ket{ib}$ represent the matrix elements of the orbital part of the one-electron spin-orbit coupling operator.
%The purple arrows labeled $\bm{\Lambda}_i=\bra{ia}\hat{\bm \Lambda}\ket{ib}$ denote the matrix elements of the orbital component of the one-electron spin–orbit coupling (SOC) operator.
(b) Schematic illustration of the orbital energy levels, the Fermi distribution functions of the electrodes, and the relevant interactions. 
$\gamma_{L/R,il\sigma}$ $(i=1,2, l = a,b;\ \sigma = \uparrow,\downarrow)$ denotes the coupling energy between orbital $il\sigma$ and the left/right electrode. 
$t_{ll}$ $(l = a,b)$ is the hopping energy between orbital $l$ of molecule 1 and molecule 2. $\hat{\bm \Lambda} \cdot \hat{s}$ represents the intramolecular spin-orbit interaction. 
$U_a$,$U_b$,$U_{ab}$ are on-site Coulomb interactions. $f_L$ and $f_R$ denote the occupation probabilities of single-particle states in the left and right electrodes, respectively.
}
\end{figure}

\section{Theory}

\subsection{Model} 
As a minimal model for a chiral molecular aggregate, we consider a dimer of two identical anisotropic molecules (molecules 1 and 2). We consider configurations in which molecule 2 is rotated by $\pm 90^\circ$ about the $z$ axis, forming a twisted structure. These configurations are related by mirror symmetry with respect to the $xz$ plane. For the current simulation, we consider a system in which the left electrode, molecule 1, molecule 2, and the right electrode are aligned in this order along the $z$ direction [Fig. \ref{fig:1}(a)].

For chiral dimer, we consider two nondegenerate molecular orbitals, $a$ and $b$, on each molecule, and express the Hamiltonian as follows: 
\begin{align}
{\hat H} &= {\hat H}_0 + {\hat H}_t + {\hat H}_{soc} + {\hat H}_{U}, \label{eq:ham}\\
{\hat H}_0 &= \sum_{i=1,2} \sum_{l=a,b} \sum_{\sigma} \epsilon_{il} {\hat c}_{il\sigma}^{\dagger} {\hat c}_{il\sigma},\\
{\hat H}_t &= \sum_{\sigma} (t_{aa} {\hat c}_{1a\sigma}^{\dagger} {\hat c}_{2a\sigma} + t_{bb} {\hat c}_{1b\sigma}^\dagger c_{2b\sigma} + h.c.),\\
{\hat H}_{soc} &= \sum_{i=1,2} \sum_{\alpha=x,y,z} i\Lambda_{i\alpha} ( {\hat c}_{ib\uparrow}^{\dagger},
{\hat c}_{ib\downarrow}^{\dagger} ) {s}_{\alpha} 
\begin{pmatrix}
 {\hat c}_{ia\uparrow}\\
 {\hat c}_{ia\downarrow} 
\end{pmatrix}, \\
{\hat H}_U &= \sum_{i=1,2} \left\{ U_{a}{\hat n}_{ia\uparrow} {\hat n}_{ia\downarrow} + U_b{\hat n}_{ib\uparrow} {\hat n}_{ib\downarrow} \right. \nonumber \\ 
&\quad \quad \quad \quad + U_{ab}^{s}\left({\hat n}_{ia\uparrow} {\hat n}_{ib\downarrow}+{\hat n}_{ia\downarrow} {\hat n}_{ib\uparrow} \right) \nonumber \\ 
&\left. \quad \quad \quad \quad + U_{ab}^{t}\left({\hat n}_{ia\uparrow} {\hat n}_{ib\uparrow} + {\hat n}_{ia\downarrow} {\hat n}_{ib\downarrow} \right) \right\}.
\end{align}
Here, ${\hat c}_{il\sigma}^\dagger$ (${\hat c}_{il\sigma}$) denotes the creation (annihilation) operator of an electron with spin $\sigma$ in orbital $l$ of monomer $i$ ($\ket{il\sigma}=\ket{il}\ket{\sigma}$). 
In ${\hat H}_{0}$, $\epsilon_{il}$ represents the energy of orbital $\ket{il}$. 
In ${\hat H}_{t}$, $t_{ll}$ ($l = a,b$) denotes the (real) hopping amplitude between $\ket{1l\sigma}$ and $\ket{2l\sigma}$. %, and hopping between $\ket{1a(b)\sigma}$ and $\ket{2b(a)\sigma}$ is neglected. 
%In ${\hat H}_{SOC}$, we consider one-electron SOC operator of the form $i \hat{\bm{\Lambda}} \cdot \bm{s}$, where $\bm{s}$ denotes the vector of Pauli matrices, 
%and define $\bm{\Lambda}_j=\bra{ja} \hat{\bm{\Lambda}} \ket{jb}$ ($j = 1,2$). 
In ${\hat H}_{SOC}$, $\Lambda_{i\alpha}=\bra{ia} \hat{\bm{\Lambda}} \ket{ib} (i=1,2)$ are the off-diagonal elements of the orbital parts of the one-electron SOC operator, and $s_\alpha$ ($\alpha=x,y,z$) denotes the Pauli matrices. 
In ${\hat H}_{U}$, $U_{ll}$ and $U_{ab}^{s/t}$ represent intra- and inter-orbital on-site Coulomb interactions, respectively, with ${\hat n}_{il\sigma} = {\hat c}_{il\sigma}^\dagger {\hat c}_{il\sigma}$.
In $H_{t}$, we neglect hopping between $\ket{1a(b)\sigma}$ and $\ket{2b(a)\sigma}$, and in $H_{U}$ we neglect pair-hopping terms such as ${\hat c}_{ia\uparrow}^\dagger {\hat c}_{ia\downarrow}^\dagger {\hat c}_{ib\downarrow}{\hat c}_{ib\uparrow}$.
Our model contains Coulomb interaction because in molecular aggregates, the transfer interaction between molecules is weak compared with the on-site Coulomb interaction.

In the following calculations, we introduce $\lambda$ as a parameter for SOC strength, and define ${\bm \Lambda}_i (i=1,2)$ as $\bm{\Lambda}_i =\lambda {\bm e}_i$, where ${\bm e}_i$ is unit vector. 
We further set $\epsilon_{1a} = \epsilon_{2a} = 0$ and $\epsilon_{1b} = \epsilon_{2b} = \varepsilon$, and assume $U_{ab}^s = U_{ab}^t = U_{ab}$, neglecting Hund's coupling.

\subsection{Multi electron current simulation} \label{subsec:IIB}
%Within the GKSL master equation formalism and under the wide-band limit approximation, 
%we neglect the energy dependence of the coupling strengths \(\gamma\) and approximate the Fermi distributions by constant values \(f_L\) and \(f_R\).
%We treat the left and right electrodes as fermionic Lindblad reservoirs and approximate the Fermi distributions by constant values $f_L$ and $f_R$.
%Under these conditions, we describe the time evolution of the multi-electron density matrix $\rho$ of the chiral dimer using the GKSL master equation \cite{PhysRevB_Jin_2020}:
We model the left and right electrodes as fermionic Lindblad reservoirs and describe the time evolution of the multi-electron density matrix $\rho$ of the chiral dimer using the GKSL master equation \cite{PhysRevB_Jin_2020}:
\begin{align}
\frac{d\rho}{dt}&=-\frac{i}{\hbar}[{\hat H},\rho]+\mathcal{L}[\rho], \\
\mathcal{L}[\rho]&=\sum_{i=1,2}\sum_{l=a,b}\sum_{\sigma=\uparrow,\downarrow} \alpha_{il\sigma} \left({\hat c}_{il\sigma}^{\dagger}\rho {\hat c}_{il\sigma}-\frac{1}{2}\left\{ {\hat c}_{il\sigma} {\hat c}_{il\sigma}^{\dagger},\rho\right\} \right) \nonumber \\
&+\sum_{i=1,2}\sum_{l=a,b}\sum_{\sigma=\uparrow,\downarrow} \beta_{il\sigma} \left({\hat c}_{il\sigma}\rho {\hat c}_{il\sigma}^{\dagger}-\frac{1}{2}\left\{ {\hat c}_{il\sigma}^{\dagger} {\hat c}_{il\sigma},\rho\right\} \right) .\label{eq:GKSL}
\end{align}
Here, $\alpha_{il\sigma}$ denote the electron injection rate from the reservoirs into the orbital $\ket{il}\ket{\sigma}$, and $\beta_{il\sigma}$ denote the electron emission rate from the orbital $\ket{il}\ket{\sigma}$ to the reservoirs. 
We define these rates as
\begin{align}
 \alpha_{il\sigma}&=\frac{\gamma_{L,il\sigma}}{\hbar}f_L+\frac{\gamma_{R,il\sigma}}{\hbar}f_R, \label{eq:alpha}\\ 
 \beta_{il\sigma}&=\frac{\gamma_{L,il\sigma}}{\hbar}(1-f_L)+\frac{\gamma_{R,il\sigma}}{\hbar}(1-f_R), \label{eq:beta}
\end{align}
where $\gamma_{L/R,il\sigma}$ describes the coupling strength between the orbitals of the left/right-electrodes and $\ket{il}\ket{\sigma}$, 
and $f_{L/R}$ denote the electron occupations (Fermi distributions) of the left/right electrodes.
%assuming energy-independent Fermi distributions for the left and right electrodes with constant values $f_L$ and $f_R$. 

Using the steady state density matrix $\rho_{ss}=\rho(\infty)$, we obtain the steady-state current $I_{tot}$ flowing from the left to the right electrode as 
\begin{align}
I_{tot}&=I_{tot,L}=-I_{tot,R}\\
I_{tot,L/R}&= \frac{q}{\hbar}\sum_{il\sigma}\gamma_{L/R,il\sigma}\Bigl\{ 
f_{L/R}\operatorname{Tr}\left[ {\hat c}_{il\sigma} {\hat c}_{il\sigma}^\dagger {\rho}_{ss} \right]\nonumber\\
&\qquad-(1-f_{L/R})\operatorname{Tr}\left[{\hat c}_{il\sigma}^{\dagger} {\hat c}_{il\sigma} \rho_{ss} \right] 
\Bigr\} \label{eq:current}.
\end{align}
Note that for a non-interacting Hamiltonian ${\hat H}$, Eq.~\eqref{eq:current} reduces to the Landauer formula \cite{Nozaki}. 

\subsection{Mean-field current simulation} \label{subsec:IIC}
To gain an intuitive understanding of the electron dynamics, we also consider the current within a mean-field approximation. %[Eq. \eqref{eq:current}] 

Within the mean-field approximation, the many-body Hamiltonian $\hat{H}$ [Eq.~\eqref{eq:GKSL}] is replaced by an effective one-electron Hamiltonian of the form
\begin{align}
\hat{H}_{\mathrm{MF}}=\sum_{n=1}^{8}\sum_{m=1}^{8} h_{nm}' \hat{c}_{n}^\dagger \hat{c}_{m},\label{eq:MFH}
\end{align}
where the operators $\hat{c}_{n}$ ($n=1,\dots,8$) are defined as 
\begin{align}
&(\hat{c}_{1},\hat{c}_{2},\hat{c}_{3},\hat{c}_{4},
\hat{c}_{5},\hat{c}_{6},\hat{c}_{7},\hat{c}_{8})\nonumber \\
&=(\hat{c}_{1a\uparrow},\hat{c}_{1b\downarrow},\hat{c}_{2a\uparrow},\hat{c}_{2b\downarrow},
\hat{c}_{1a\downarrow},\hat{c}_{1b\uparrow},\hat{c}_{2a\downarrow},\hat{c}_{2b\uparrow}). 
\end{align}
%and the coefficients $h'_{nm}$ denote the matrix elements of the effective one-electron Hamiltonian matrix $\bm{h}'$. 
We construct effective one-electron Hamiltonian matrix $\bm{h}'$ with elements $h'_{nm}$ as follows:
\begin{align}
\bm{h}'=\bm{h}+{\bm \Delta}({\bm n}). \label{eq:hprime}
\end{align}
Here, $\bm{h}$ represents one-electron hamiltonian matrix corresponding to $\hat{H}_{0}+\hat{H}_{t}+\hat{H}_{\mathrm{SOC}}$ [Eq. \eqref{eq:ham}], while ${\bm \Delta}({\bm n})$ describes the mean-field correction derived from on-site Coulomb interaction (${\hat H}_{U}$). 
The matrix elements of ${\bm \Delta}({\bm n})$ are given by %\footnote{${\bar \sigma}$ denotes the spin opposite to $\sigma$}
\begin{align}
&\Delta_{il\sigma,i'l'\sigma}({\bm n}) \nonumber \\
&=
\delta_{i i'}\delta_{l l'}\delta_{\sigma \sigma'}
\left(
U_{l}n_{il\bar \sigma,il\bar \sigma} 
+U_{ab}\left(n_{i{\bar l}\uparrow,i{\bar l}\uparrow}+n_{i{\bar l}\downarrow,i{\bar l}\downarrow} \right)
\right)\nonumber\\
&\quad+\delta_{i i'}\delta_{l l'}(1-\delta_{\sigma \sigma'})
U_{l}n_{il\sigma',il\sigma} \nonumber \\
&\quad+\delta_{i i'}(1-\delta_{l l'})(1-\delta_{\sigma \sigma'})
U_{ab}n_{il'\sigma',il\sigma},
\label{eq:Delta}
\end{align}
where  ${\bar \sigma}$ denotes the spin opposite to $\sigma$ (${\bar \uparrow}=\downarrow, {\bar \downarrow}=\uparrow$), and ${\bar l}$ does the other orbital (${\bar a}=b, {\bar b}=a$). 
$\bm{h}'$ should be determined self-consistently so that Eq.~\eqref{eq:hprime} satisfied for ${\bm n}$ with elements
\begin{align}
n_{il\sigma,i'l'\sigma'}=\mathrm{Tr}\left[\hat c_{il\sigma}^\dagger \hat c_{i'l'\sigma'}\rho_{\mathrm{ss}}\right],
\end{align}
where $\rho_{ss}$ is the steady state density matrix obtained with ${\bm h}'$. 
Once ${\bm h}'$ is determined, the cureent can be evaluated using Eq. \eqref{eq:current}.

\subsection{{MR-CISS}}
In this work, we define spin polarization in MR-CISS as
\begin{align}
P_{MR}= 
\frac{I_{\mathrm{tot}}^{\uparrow} - I_{\mathrm{tot}}^{\downarrow}}{I_{\mathrm{tot}}^{\uparrow}+I_{\mathrm{tot}}^{\downarrow}}.
\end{align}
Here, $I_{\mathrm{tot}}^{\sigma'}$ ($\sigma' = \uparrow, \downarrow$) denotes the current flowing from a spin-$\sigma'$ polarized left electrode to a spin-unpolarized right electrode. 

We denote coupling strengths between the spin-$\sigma'$-polarized left electrode and the molecular orbital $\ket{il\sigma}$ as $\gamma_{L,il\sigma}^{\sigma'}$, 
and that between the spin-unpolarized right electrode and $\ket{il\sigma}$ as $\gamma_{R,il\sigma}$. 
As summarized in Table \ref{tab:1}, we retain only spin-conserving couplings between adjacent molecules and electrodes; namely, we set 
$\gamma_{L,2l\uparrow}^{\uparrow}=\gamma_{L,2l\downarrow}^{\downarrow}=\gamma_{L,1l\uparrow}^{\downarrow}=\gamma_{L,1l\downarrow}^{\uparrow}=\gamma_{R,1l\uparrow}=\gamma_{R,1l\downarrow}=0$ ($l=a,b$). 
Additionally, we assume $\gamma_{L,1l\sigma}^{\sigma}=\gamma_{R,2l\uparrow}=\gamma_{R,2l\downarrow}=\gamma_l$ to reduce the number of parameters.

By substituting the parameters $\gamma_{L,il\sigma}^{\sigma'}$ and $\gamma_{R,il\sigma}$, listed in Table~\ref{tab:1}, into $\gamma_{L,il\sigma}$ and $\gamma_{R,il\sigma}$ in Eqs.~\eqref{eq:alpha} and \eqref{eq:beta}, 
we obtain $I_{\mathrm{tot}}^{\sigma'}$ as 
\begin{align}
I_{\mathrm{tot}}^{\sigma'}&= \frac{q}{\hbar}\sum_{l=a,b} \gamma_l \left(f_L - n_{1a\sigma',1a\sigma'}^{\sigma'} \right)\nonumber \\
&= \frac{q}{\hbar}\sum_{\sigma=\uparrow,\downarrow} \sum_{l=a,b} \gamma_l \left( n_{2l\sigma,2l\sigma}^{\sigma'} - f_R \right). \label{eq:MRcurrent}
\end{align}
Here, we define 
\begin{align}
n_{il\sigma,i'l'\sigma'}^{\sigma'}=\mathrm{Tr}\left[\hat c_{il\sigma}^\dagger \hat c_{i'l'\sigma'}\rho_{ss}^{\sigma'}\right], 
\end{align}
where the superscript $\sigma'$ in $n_{il\sigma,i'l'\sigma'}^{\sigma'}$ and in the steady state density matrix $\rho_{ss}^{\sigma'}$ indicates the spin polarization of the left electrode.
%and $n_{il\sigma,i'l'\sigma'}^{\sigma'}$ denotes the one-electron density matrix of the steady state $\rho_{\mathrm{ss}}^{\sigma'}$, given by

%As shown in Table~\ref{tab:1}, we consider only the coupling between the left (right) electrode and molecule 1 (2). Additionally, we assume $\gamma_{L,1l\sigma}^{\sigma}=\gamma_{R,2l\uparrow}=\gamma_{R,2l\downarrow}=\gamma_l$ to reduce the number of parameters.

\begin{table}[htbp]
\caption{Coupling strengths between the spin-$\sigma'$-polarized left electrode and the molecular orbital $\ket{il\sigma}$ ($\gamma_{L,il\sigma}^{\sigma'}$), 
and between the spin-unpolarized right electrode and $\ket{il\sigma}$ ($\gamma_{R,il\sigma}$). }
\begin{ruledtabular} 
\label{tab:1}
\begin{tabular}{ccccccccc} 
$il\sigma$&$1a\uparrow$&$1b\downarrow$&$2a\uparrow$&$2b\downarrow$&$1a\downarrow$&$1b\uparrow$&$2a\downarrow$&$2b\uparrow$\\\hline
${\gamma}_{L,il\sigma}^\uparrow$&$\gamma_a$&0&0&0&0&$\gamma_b$&0&0\\    ${\gamma}_{L,il\sigma}^\downarrow$&0&$\gamma_b$&0&0&$\gamma_a$&0&0&0\\${\gamma}_{R,il\sigma}$&0&0&$\gamma_a$&$\gamma_b$&0&0&$\gamma_a$&$\gamma_b$\\
\end{tabular}
\end{ruledtabular}
\end{table}

\section{Results and Discussion} 
We investigated MR-CISS based on multi-electron current simulations. 
As proved in Appendix \ref{appendix}, MR-CISS does not occur for a non-interacting system Hamiltonian. 
Therefore, we focused on the interacting Hamiltonian. 
All calculations were performed with ${\bm \Lambda}_{1}=(\lambda,0,0)$, ${\bm \Lambda}_{2}=(0,\lambda,0)$, $t_{aa}=t$ and $\lambda=0.03t$.  
Note that $P_{MR}$ calculated with ${\bm \Lambda}_{1} = (\lambda,0,0)$ and ${\bm \Lambda}_{2} = (0,\pm\lambda,0)$ takes opposite signs, indicating that its sign is determined by chirality of the system.

\subsection{{Fermi distribution dependence of MR-CISS}}\label{subsec:fL} 
We first examined how nonequilibrium conditions affect transport and the emergence of MR-CISS, 
using $|f_a - 0.5|$ ($a = L, R$) as a measure of nonequilibrium. 
We varied $f_L$ and $f_R$ under the constraint $f_R= 1 - f_L$ for $\varepsilon=0.4t,0.6t,0.8t,1.0t,1.2t,1.4t$, 
and set the other parameters to $\gamma_a = \gamma_b = 0.1t$, 
$U_a = U_b = U_{ab} = 4t$, $t_{aa} = t$, $t_{bb} = 0.2t$, and $\lambda = 0.03t$. 

\begin{figure}[htbp]
\centering
\begin{tabular}{l}
\includegraphics[bb=0 0 576 1008, width=0.9\linewidth]{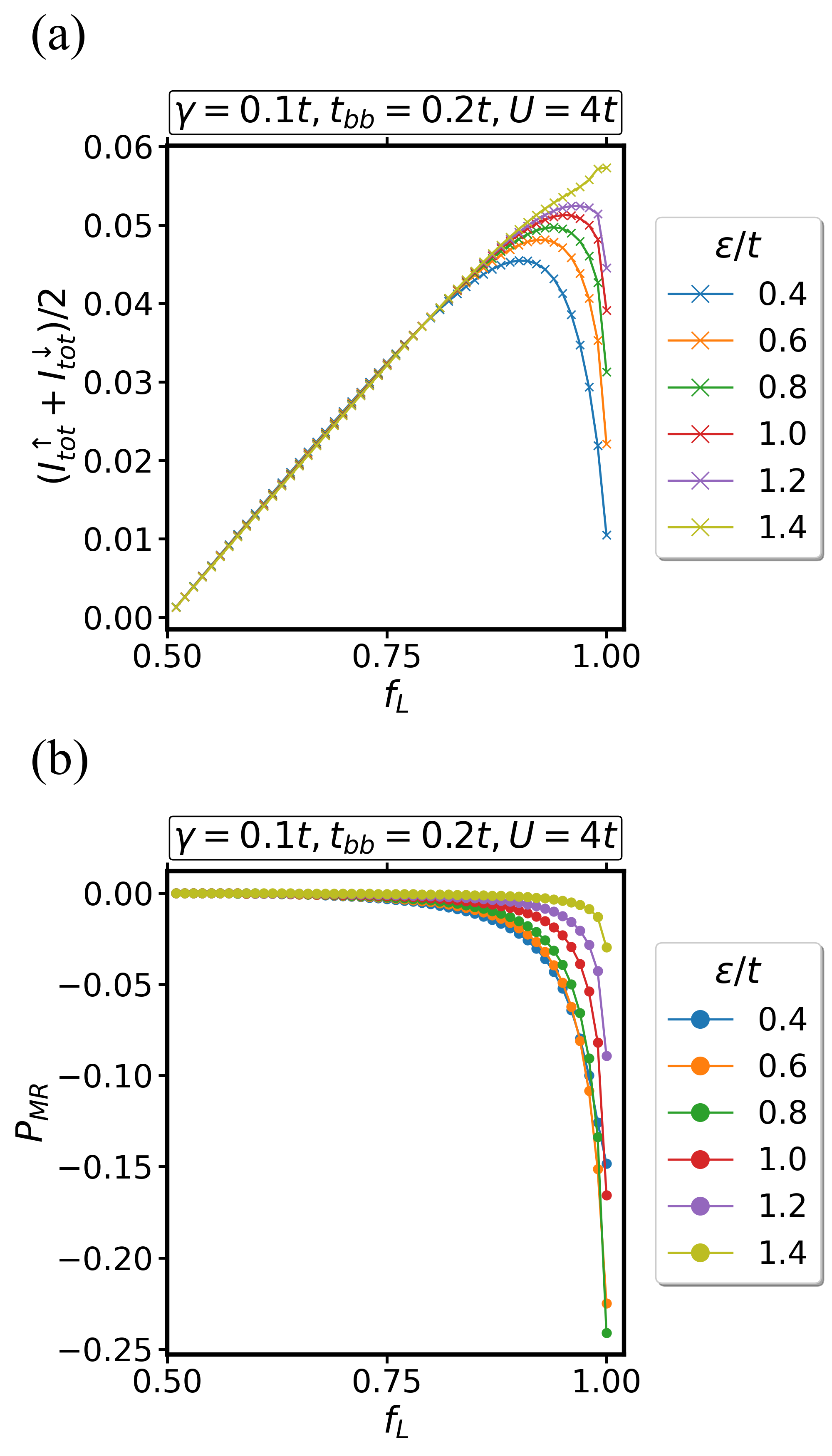} %
\end{tabular}
\caption{
$f_L$ dependence of the average currents $(I_{tot}^{\uparrow} +I_{tot}^{\downarrow})/2$ 
(a) and the spin polarization $P_{MR}$ (b) with  $f_R=1-f_L$, $U=U_{a}=U_{b}=U_{ab}$, $t_{bb}=0.2t$, $\gamma_a=\gamma_b=0.1t$, $t_{aa}=t$ and $\lambda=0.03t$. 
In (a) and (b), blue, orange, green, red, purple, and olive markers correspond to  
$\varepsilon=0.4t,0.6t,0.8t,1.0t,1.2t,1.4t$. 
}
\label{fig:4}
\end{figure}

As shown in Fig.~\ref{fig:4}(a), for all $\varepsilon$, average current $(I_{tot}^\uparrow+I^\downarrow_{tot})/2$ increases linearly with $f_L$ in the low-$f_L$ regime, while in the high-$f_L$ regime it deviates from linearity and decreases.  
In contrast, in the non-interacting case ($U=0$), the current remains linear in $f_L$ over the entire range (data not shown). This indicate that the observed nonlinear behavior originates from the on-site Coulomb interaction. 
Additionally, at $f_L = 1$, the current becomes smaller as $\varepsilon$ decreases. 
%the current is smaller for the smaller values of $\varepsilon$. 
This behavior can be attributed to the enhanced effect of inter-orbital Coulomb interactions for smaller energy gaps. 
%This behavior can be attributed to an amplified effective inter-orbital Coulomb interaction for smaller energy gaps. 

As shown in Fig.~\ref{fig:4}(b), for all $\varepsilon$, magnitude of $P_{MR}$ is nearly zero in the low-$f_L$ regime, while in the high-$f_L$ regime it rapidly increases along with the decrease of average current [Fig.~\ref{fig:4}(a)]. This correlation indicates that the emergence of MR-CISS is closely related to nonlinear electron transport. 
In contrast to the average current, the magnitude of $P_{MR}$ at $f_L = 1$ does not show a monotonic dependence with $\varepsilon$. 
This behavior is somewhat counterintuitive, given that both effective spin-orbit coupling and on-site Coulomb interactions are amplified for smaller energy gaps. To gain further insight into this behavior, we investigated the parameter dependence of $P_{MR}$ in Sec.~\ref{subsec:Utg}.

\subsection{{Dependence of MR-CISS on model parameters}}\label{subsec:Utg}
Next, we investigated the conditions under which MR-CISS is amplified, assuming $U_a = U_b = U_{ab} = U$ and $\gamma_a = \gamma_b = \gamma$.
In accordance with Sec.~\ref{subsec:fL}, we focused on the strongly nonequilibrium regime by setting $f_L = 1$ and $f_R = 0$.
We chose $(U, t_{bb}, \gamma) = (4t, 0.2t, 0.1t)$ as a reference parameter set and varied each parameter independently while keeping the other two fixed.
These parameter scans were performed for $\varepsilon = 0.4t, 0.6t, 0.8t, 1.0t, 1.2t,$ and $1.4t$.

\begin{figure*}[htbp]
\centering
\begin{tabular}{l}
\includegraphics[bb=0 0 1465 932, width=0.99\linewidth, valign=c]{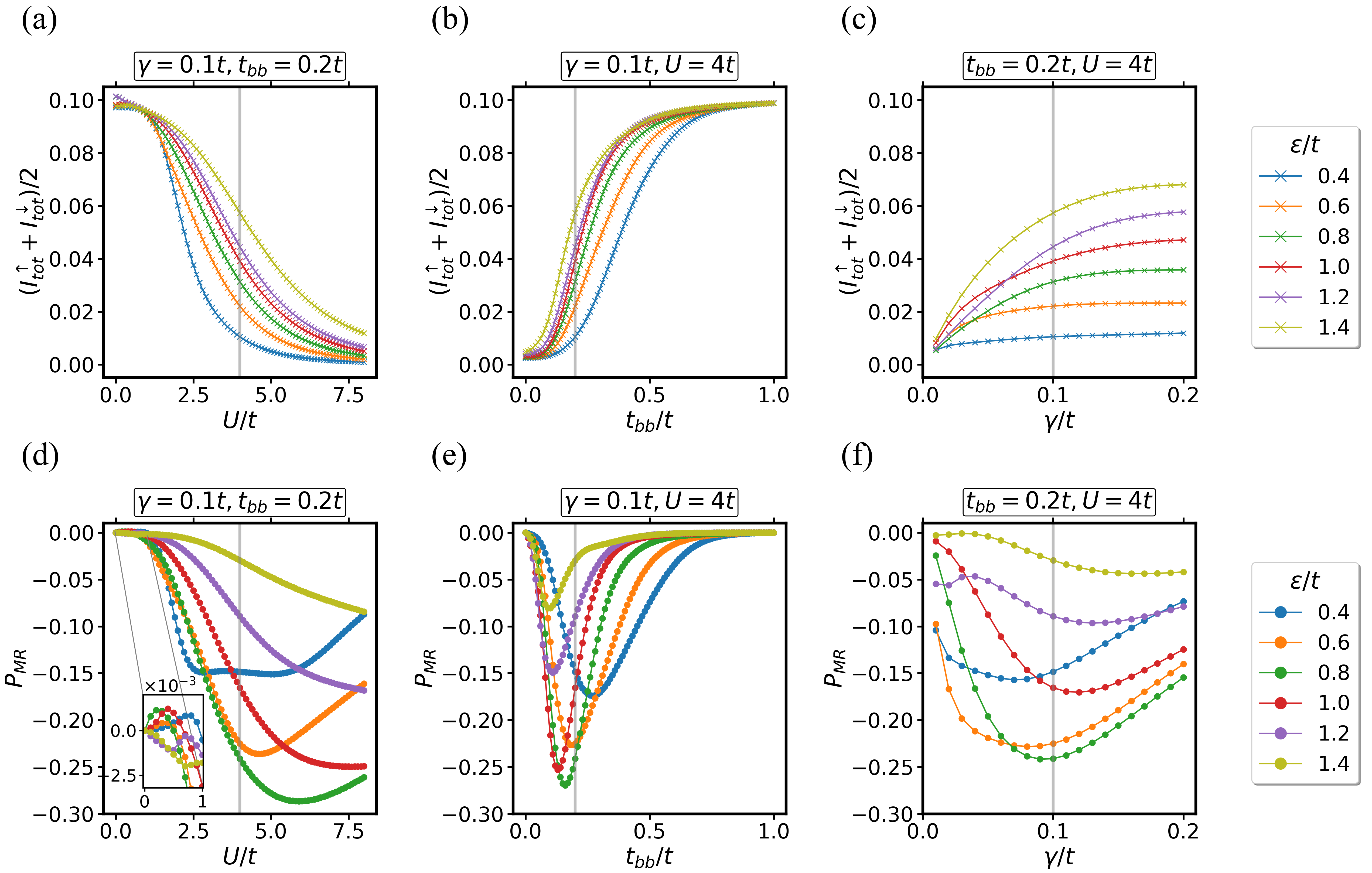} 
\end{tabular}
\caption{
Dependence of the average total current $(I_{\mathrm{tot}}^{\uparrow} + I_{\mathrm{tot}}^{\downarrow})/2$ [(a)-(c)] and the spin polarization $P_{MR}$ [(d)-(f)] on $U(=U_a=U_b=U_{ab})$ [(a),(d)], $t_{bb}$ [(b),(e)], and $\gamma(=\gamma_a=\gamma_b)$ [(c),(f)], with $f_L=1$, $f_R=0$, $\lambda=0.03t$, and $t_{aa}=t$.
In (a) and (d), $t_{bb}=0.2t$ and $\gamma_a=\gamma_b=0.1t$ are fixed.
In (b) and (e), $U_a=U_b=U_{ab}=4t$ and $\gamma_a=\gamma_b=0.1t$ are fixed.
In (c) and (f), $U_a=U_b=U_{ab}=4t$ and $t_{bb}=0.2t$ are fixed.
The gray vertical lines indicate $(U,t_{bb},\gamma)=(4t,0.2t,0.1t)$.
The color mapping for $\varepsilon$ is identical to that in Fig.~\ref{fig:4}.}
\label{fig:2}
\end{figure*}

The average current exhibits distinct nonlinear dependencies on $U$, $t_{bb}$, and $\gamma$ [Figs.~\ref{fig:2}(a)-(c)].
Overall, it increases with decreasing $U$ [Fig.~\ref{fig:2}(a)] and with increasing $t_{bb}$ and $\gamma$ [Figs.~\ref{fig:2}(b) and (c)] for all values of $\varepsilon$ we considered.
A comparison of the average current calculated for the same values of $(U, t_{bb}, \gamma)$ but different $\varepsilon$ shows that the current increases as $\varepsilon$ decreases. 
As discussed in Sec.~\ref{subsec:fL}, this trend can be attributed to the enhanced role of effective on-site Coulomb interactions at smaller energy gaps. 
%this trend can be attributed to the amplification of the effective on-site Coulomb interaction for smaller energy gaps.

For the $U$ scan of $\varepsilon = 0.4t, 0.6t,$ and $0.8t$ [Fig.~\ref{fig:2}(d)], $P_{MR}$ exhibits a minima. 
The minima position shifts to larger values of $U$ as $\varepsilon$ increases.
Based on this trend, we expect that $P_{MR}$ for $\varepsilon = 1.2t$ and $1.4t$ also develops a minima outside the range of $U$ investigated here, although only a monotonic increase is observed in Fig.~\ref{fig:2}(d). 
%In Fig.~\ref{fig:2}(d), the largest magnitude of $P_{MR}$ is found at $U \approx 6.0t$ and $\varepsilon = 0.8t$, where $P_{MR} = -0.29$.
In Fig.~\ref{fig:2}(d), $|P_{MR}|$ is largest at $U=6.0t$ and $\varepsilon = 0.8t$, where $P_{MR} = -0.29$.

For the $t_{bb}$ scan, $P_{MR}$ exhibits a clear resonance as a function of $t_{bb}$ for all values of $\varepsilon$ considered [Fig.~\ref{fig:2}(e)].
The peak position roughly corresponds to the value of $t_{bb}$ at which the average current reaches 0.02 for $\gamma = 0.1t$ and $U = 4t$. 
Furthermore, $P_{MR}$ vanishes at $t_{bb} = 0$ and $t_{bb} = t$.
%In Fig.~\ref{fig:2}(e), the largest magnitude of $P_{MR}$ is found at $t_{bb} = 0.16t$ and $\varepsilon = 0.8t$, where $P_{MR} = -0.26$.
In Fig.~\ref{fig:2}(e), $|P_{MR}|$ is largest at $t_{bb} = 0.16t$ and $\varepsilon = 0.8t$, where $P_{MR} = -0.26$.

For the $\gamma$ scan, $P_{MR}$ increases with $\gamma$, reaches a peak, and then decreases for $\varepsilon = 0.4t, 0.6t,$ and $0.8t$ [Fig.~\ref{fig:2}(f)].
For larger $\varepsilon$ ($\varepsilon = 1.2t$ and $1.4t$), $P_{MR}$ shows only a weak dependence on $\gamma$, with two local maxima and one local minimum.
%In Fig.~\ref{fig:2}(f), The largest magnitude of $P_{MR}$ is obtained at $\gamma = 0.09t$ and $\varepsilon = 0.8t$, where $P_{MR} = -0.24$.
In Fig.~\ref{fig:2}(f), $|P_{MR}|$ is largest at $\gamma = 0.09t$ and $\varepsilon = 0.8t$, where $P_{MR} = -0.24$.

To summarize these results, large $|P_{MR}|$ is favored in the regime $U \gg t_{aa}$ and $t_{aa} \gg t_{bb}$, particularly for large $\varepsilon$.  
%In summary, the regime $U \gg t_{aa}$ and $t_{aa} \gg t_{bb}$ is optimal for achieving a large $|P_{MR}|$, especially with high values of $\varepsilon$.
Our results also suggest that excessively small $\gamma$, excessively small $t_{bb}$, or excessively large $U$ do not amplify MR-CISS. %taking $(U, t_{bb}, \gamma) = (4t, 0.2t, 0.1t)$ as a reference,

\subsection{{Mechanism of MR-CISS}}\label{subsec:mechanism} 
In Secs.~\ref{subsec:fL} and \ref{subsec:Utg}, we confirmed the emergence of MR-CISS under the nonequilibrium conditions in the presence of intra-site SOC, on-site Coulomb interaction, and asymmetric hopping ($t_{aa} \neq t_{bb} \neq 0$).  

To understand how MR-CISS emerges within our model, we considered six distinct cases [cases (a)-(f)], as summarized in Table~\ref{tab:1}. We fixed the parameters at $t_{bb} = 0.2t$, $\varepsilon = 0.8t$, $f_L = 1$ and $f_R = 0$, while varied $U_{a}$, $U_{b}$, $\gamma_a$, and $\gamma_b$ among the six cases.
In case (f), we set $U_{a} = U_{b} = U_{ab} = 4t$ and $\gamma_a = \gamma_b = 0.1t$, wchich are the reference parameter used in Sec.~\ref{subsec:Utg}. For this case, we obtained $P_{MR} = 0.24$.

First, we consider the case with $\gamma_a = 0.1t$ and $\gamma_b = 0$ [Table~\ref{tab:2}(a),(c)].
In this case, electrons injected from a $\sigma$-spin-polarized left electrode can transition only to the states $\ket{1a\sigma}$, $\ket{1b\bar{\sigma}}$, $\ket{2a\sigma}$, and $\ket{2b\bar{\sigma}}$.
As shown in Table \ref{tab:2}(a) and (b), we obtained $P_{MR}=0$ for $U_{ab}=0$ and $U_{a}=U_{b}=4t$ [case (a)], and $P_{MR}=-0.0031$ for $U_{ab}=4t$ and $U_{a}=U_{b}=0$ [case (b)].  
Interestingly, the value of $P_{MR}$ for $U_{ab}=4t$ and $U_{a}=U_{b}=4t$ [case (c)] is the same as that for $U_{ab}=4t$ and $U_{a}=U_{b}=0$ [case (b)].

These results can be understood within the mean-field approximation described in Sec.~\ref{subsec:IIC}.
In case (a), $U_l$ ($l = a,b$) do not contribute to ${\bm \Delta}({\bm n}^\sigma)$ defined in Eq.~\eqref{eq:Delta}, because $n_{n,m}^\sigma = 0$ for $(n,m \in \{1a\bar{\sigma}, 1b\sigma, 2a\bar{\sigma}, 2b\sigma\})$.
Consequently, $P_{MR} = 0$ follows from ${\bm \Delta}({\bm n}^{\uparrow}) = {\bm \Delta}({\bm n}^{\downarrow}) = 0$.
In contrast, in cases (b) and (c), $U_{ab}$ contributes to ${\bm \Delta}({\bm n}^\sigma)$ through $n_{n,m}^\sigma$ $(n,m \in \{1a\sigma, 1b\bar{\sigma}, 2a\sigma, 2b\bar{\sigma}\})$.
Moreover, ${\bm \Delta}({\bm n}^{\uparrow}) \neq {\bm \Delta}({\bm n}^{\downarrow})$ is ensured by
\begin{align}
n_{1b\downarrow,1b\downarrow}^\uparrow \neq n_{1b\uparrow,1b\uparrow}^\downarrow.
\label{eq:nn}
\end{align}
As a consequence, $P_{MR} \neq 0$ in cases (b) and (c).
Equation~\eqref{eq:nn} holds even in the noninteracting case $U_a = U_b = U_{ab} = 0$, as it originates from the spin-orbit interaction.

Next, we consider the case with $\gamma_{a}=0.1t$ and $\gamma_{b}=0.1t$  [Table \ref{tab:2}(d)-(f)]. 
As shown in Table \ref{tab:2}(d) and (e),
We obtained $P_{MR}=0.013$ for $U_{a}=U_{b}=4t$ and $U_{ab}=0$ [case (d)], and $P_{MR}=0.012$ for $U_{a}=U_{b}=0$ and $U_{ab}=4t$ [case (e)]. 
In contrast to case (a), we obtained $P_{MR}\neq0$ in case (d) because of $n_{n,m}^\sigma\neq0$ $(n,m \in \{1a{\bar \sigma}, 1b\sigma, 2a\bar{\sigma},2b\sigma\})$. 
Compared with case (f), $P_{MR}$ in cases (d) and (e) are significantly small. These results indicate a constructive interplay between inter-orbital and intra-orbital onsite Coulomb interactions within the MR-CISS mechanism.

\begin{table}[htbp]
\caption{Evaluation of $P_{MR}$ for different values of ($\gamma_b$,$U_{ab}$,$U_{a}$,$U_{b}$) with $f_L=1$, $f_R=0$, $t_{aa}=t$, $t_{bb}=0.2t$, $\varepsilon=0.8t$, $\lambda=0.03t$ and $\gamma_{a}=0.1t$.}
\label{tab:2}
\begin{ruledtabular} 
\begin{tabular}{ccccc}
&$\gamma_{b}$ & $U_{ab}$ & $U_{a},U_{b}$ &  $P_{MR}(\times100)$\\
\hline
(a) & 0 & 0 & 4t & 0   \\
(b) & 0 & 4t & 0 & -0.31 \\
(c) & 0 & 4t & 4t & -0.31 \\
(d) & 0.1t & 0 & 4t & 1.3   \\
(e) & 0.1t & 4t & 0 & 1.2 \\
(f) & 0.1t & 4t & 4t & 24 \\
\end{tabular}
\end{ruledtabular}
\end{table}

\subsection{{MR-CISS within mean-field approximation}}\label{subsec:MF}
\begin{figure}[tb]
\centering
\begin{tabular}{l}
\includegraphics[bb=0 0 576 576, width=\linewidth]{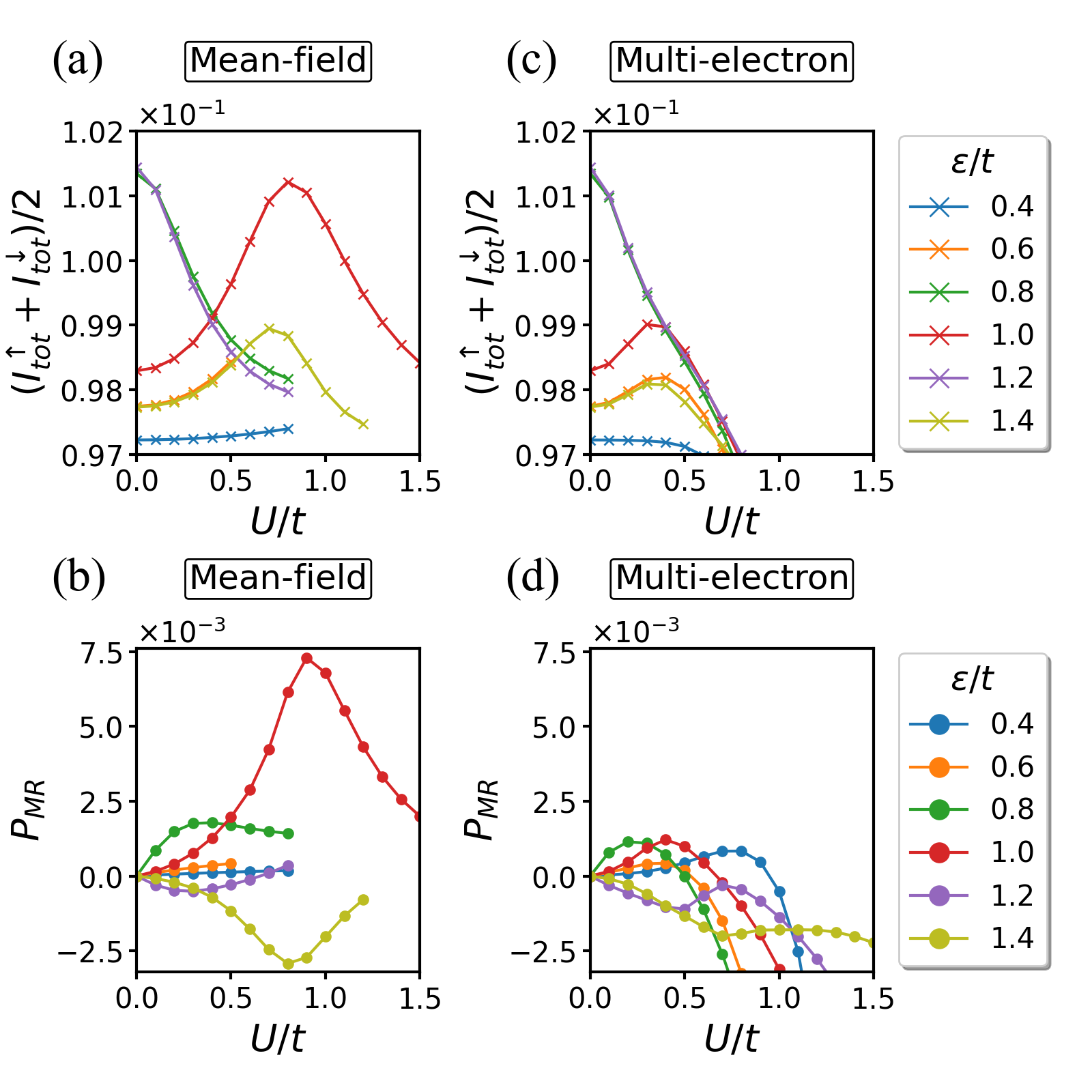}
\end{tabular}
\caption{
$U$ dependence of average currents $(I_{\mathrm{tot}}^{\uparrow} +I_{\mathrm{tot}}^{\downarrow})/2$ 
(a,c) and spin polarizations $P_{MR}$ (b,d), calculated within the mean-field approximation (a,b) and the multi-electron GKSL master equation (c,d). In (a)--(d), the parameters are set to $t_{aa}=t$, $\lambda=0.03t$, $t_{bb}=0.2t$, $\gamma_a=\gamma_b=0.1t$, $f_L=1$, $f_R=0$ and $U=U_{a}=U_{b}=U_{ab}$. 
The color mapping for $\varepsilon$ is identical to that in Fig.~\ref{fig:4}.
}
\label{fig:3}
\end{figure}

Finally, we examined how well mean-field calculations reproduce the $U$ dependence of the average current $(I_{\mathrm{tot}}^{\uparrow} + I_{\mathrm{tot}}^{\downarrow})/2$ and spin polarization $P_{MR}$ shown in Figs.~\ref{fig:2}(a) and (d). 
%For comparison, we show the same quantities as Figs.~\ref{fig:2}(a) and (d) in Figs.~\ref{fig:3}(c) and (d) respectively, but with different plotting ranges. 
For comparison, the same quantities as shown in Figs.~\ref{fig:2}(a) and (d) are presented in Figs.~\ref{fig:3}(c) and (d), respectively, but with different plotting ranges.
The mean-field calculations were performed under the same conditions as those used in Figs.~\ref{fig:2}(a) and (d). 

As shown in Figs.~\ref{fig:3}(a) and (b), for small $U$ ($U \lesssim 0.5t$), the mean-field results well reproduce the multi-electron results [Figs.~\ref{fig:3}(c) and (d)].
However, the mean-field results deviate from the multi-electron results as $U$ increases.
Additionally, the self-consistent calculations converge only for small $U$.
These results indicate that a simple mean-field treatment is insufficient to reproduce large $P_{MR}$, highlighting the importance of approaches beyond the mean-field approximation.

\section{Conclusion}

In this study, we investigated MR-CISS in a chiral dimer model using the GKSL master equation, 
focusing on the interplay between intra-site SOC and on-site Coulomb interactions.
We showed that MR-CISS is significantly enhanced under strongly nonequilibrium conditions, 
and that sizable spin polarization can emerge from the cooperative effects of intra-site SOC and on-site Coulomb interactions. 
In particular, our parameter analysis identified regimes in which $P_{MR}$ exceeds 25\%. 
Furthermore, we found that MR-CISS is strongly enhanced when both intra-orbital and inter-orbital on-site Coulomb interactions are taken into account.
We also demonstrated that the large $P_{MR}$ observed within the GKSL framework are not reprodueced by the mean-field approximation. 

The present model serves as a prototypical framework for MR-CISS in chiral molecular aggregates and provides a foundation for a quantitative understanding of the CISS effect.

%In this study, we investigated MR-CISS in a chiral dimer model using the GKSL master equation, focusing on the interplay between intra-site SOC and on-site Coulomb interactions.
%We showed that MR-CISS is significantly amplified under strongly nonequilibrium conditions, and that sizable spin polarization can arise through the cooperative effects of intra-site spin-orbit coupling and on-site Coulomb interactions.
%In particular, our parameter analysis revealed regimes in which $P_{MR}$ exceeds 25\%.
%Furthermore, we reveal that MR-CISS is strongly enhanced when both intra-orbital and inter-orbital on-site Coulomb interactions are included.
%We also demonstrated that the large amplification of $P_{MR}$ with GKSL master equation does not arise in the mean-field approximation..
%The present model serves as a prototypical framework for MR-CISS in chiral molecular aggregates and provides a basis for a quantitative understanding of CISS. 

\begin{acknowledgments}
This work was supported by Grant-in-Aid for Transformative Research Areas ``Materials Science of Meso-Hierarchy'' (No.JP23H04879). The calculations were performed using the Research Center for Computational Science, Okazaki, Japan (Project:25-IMS-C125).
\end{acknowledgments}

%\clearpage

\appendix
\section{Absence of MR-CISS in the non-interacting system}\label{appendix} 
Here, we consider $I_{tot}^\sigma$ defined in Eq. \eqref{eq:MRcurrent}, and show that $I_{\mathrm{tot}}^\uparrow = I_{\mathrm{tot}}^\downarrow$ for a non-interacting systmes ($U_a = U_b = U_{ab}^s =U_{ab}^t = 0$).  
In this proof, we express $I_{tot}^\sigma$ in the form of the Landauer formula \cite{Nozaki}:
\begin{align}
I^{\sigma}_{tot}\nonumber&=(f_{L}-f_{R}) \int_{-\infty}^{\infty} d\epsilon \Tr\left[{\bm \gamma}_{L}^{\sigma} \bm{G}^{r\sigma}(\epsilon) {\bm \gamma}_R \bm{G}^{a\sigma}(\epsilon)  \right] \nonumber \\
&=(f_{L}-f_{R}) \int_{-\infty}^{\infty} d\epsilon \Tr\left[{\bm \gamma}_{R} \bm{G}^{r\sigma} (\epsilon) {\bm \gamma}_L^{\sigma} \bm{G}^{a\sigma} (\epsilon) \right] \label{eq:landauer}
\end{align}
with
\begin{align}
&{\bm G}^{r\sigma}(\epsilon)=\left(\epsilon-{\bm h}+ \frac{i({\bm \gamma}_L^{\sigma}+{\bm \gamma}_R)}{2} +i\eta \right)^{-1}, \label{eq:12}\\
&{\bm G}^{a\sigma}(\epsilon)=\left(\epsilon-{\bm h}-\frac{i({\bm \gamma}_L^{\sigma}+{\bm \gamma}_R)}{2} -i\eta \right)^{-1}. \label{eq:13}
\end{align} 
where $\bm h$ is one electron hamiltonian matrix given in Eq. \eqref{eq:hprime}. 
${\bm \gamma}_L^{\sigma}$ and ${\bm \gamma}_R$ are diagonal matrices with elements ${\gamma}_{L,il\sigma',il\sigma'}^{\sigma}=\delta_{il\sigma',il\sigma'}{\gamma}_{L,il\sigma'}^{\sigma}$ and ${\gamma}_{R,il\sigma,il\sigma}=\delta_{il\sigma,il\sigma}\gamma_{R,il\sigma}$, respectively.

By substituting $\gamma_{L,il\sigma}^{\sigma'}$ and $\gamma_{R,il\sigma}$ given in Table \ref{tab:1} into \eqref{eq:landauer}, we obtain
\begin{align}
I^{\uparrow}_{tot}\nonumber&=(f_{L}-f_{R}) \int_{-\infty}^{\infty} d\epsilon \Tr\left[{\bm \gamma}_{L}^{\uparrow} \bm{G}^{r\uparrow}(\epsilon) {\bm \gamma}_R \bm{G}^{a\uparrow}  (\epsilon)\right] \nonumber \\
&= (f_{L}-f_{R})\sum_{l,l'=a,b}\sum_{\sigma=\uparrow,\downarrow} \gamma_l \gamma_{l'} \int_{-\infty}^{\infty} d\epsilon \left |G^{r\uparrow}_{1l\uparrow,2l'\sigma} (\epsilon) \right |^2 \nonumber \\
&= (f_{L}-f_{R})\sum_{l,l'=a,b}\sum_{\sigma=\uparrow,\downarrow} \gamma_l \gamma_{l'} \int_{-\infty}^{\infty} d\epsilon \left |G^{r\downarrow}_{2l'{\bar \sigma},1l\downarrow}(\epsilon)  \right |^2 \nonumber \\
&= (f_{L}-f_{R})\sum_{l,l'=a,b}\sum_{\sigma=\uparrow,\downarrow} \gamma_l \gamma_{l'} \int_{-\infty}^{\infty} d\epsilon \left |G^{r\downarrow}_{2l'\sigma,1l\downarrow}(\epsilon) \right |^2 \nonumber \\
&=(f_{L}-f_{R})\int_{-\infty}^{\infty} d\epsilon \Tr\left[{\bm \gamma}_{R}\bm{G}^{r\downarrow}(\epsilon) {\bm \gamma}_L^{\downarrow} \bm{G}^{a\downarrow} (\epsilon) \right] \nonumber \\
&=I^{\downarrow}_{tot} \label{eq:proof}
\end{align}
From the second to the third line in Eq. \eqref{eq:proof}, we use 
\begin{eqnarray}
 \left |G_{il\sigma,i'l'\sigma'}^{\uparrow r} \right |=\left |G^{\downarrow r}_{i'l'\bar{\sigma}',il\bar{\sigma} }\right |.
 \label{eq:lemma}
\end{eqnarray}
We show the proof of  Eq. \eqref{eq:lemma} in Appendix \ref{sec:lemma}.

\section{{Proof of Eq. \eqref{eq:lemma}}}\label{sec:lemma}
We prove that Eq. \eqref{eq:lemma} holds for 
\begin{align}
&{\bm \gamma}_L^{\sigma}=\operatorname{diag}(\gamma_a\delta_{\sigma\uparrow}, \gamma_b\delta_{\sigma\downarrow},0,0,\gamma_a\delta_{\sigma\downarrow},\gamma_b\delta_{\sigma\uparrow},0,0),\\
&{\bm \gamma}_R=\operatorname{diag}(0,0,\gamma_a,\gamma_b,0,0,\gamma_a,\gamma_b),
\end{align}
and
\begin{align}
{\bm h}&=
\begin{pmatrix}
{\bm h}_{11}&0\\
0&{\bm h}_{22}
\end{pmatrix},
\end{align}
where 
\begin{align}
{\bm h}_{11}&=
\begin{pmatrix}
e_{1a}&i\lambda e^{i\theta_1}&t_{aa}&0\\
-i\lambda e^{-i\theta_1}&e_{1b}&0&t_{bb}\\
t_{aa}&0&e_{2a}&ie^{i\theta_2}\\
0&t_{bb}&-i\lambda e^{-i\theta_2}&e_{2b}
\end{pmatrix}, \label{eq:h11} \\
{\bm h}_{22}&=
\begin{pmatrix}
e_{1a}&i\lambda e^{-i\theta_1}&t_{aa}&0\\
-i\lambda e^{i\theta_1}&e_{1b}&0&t_{bb}\\
t_{aa}&0&e_{2a}&ie^{-i\theta_2}\\
0&t_{bb}&-i\lambda e^{i\theta_2}&e_{2b} 
\end{pmatrix}.\label{eq:h22}
\end{align}
Here, we assume ${\bm \Lambda}_i = (\cos\theta_i, \sin\theta_i, 0)$ $(i=1,2)$ in $\bm h_{11}$ and $\bm h_{22}$.  
Under the unitary transformation $\bm X$, which swaps the spin labels 
($\uparrow \leftrightarrow \downarrow$), 
$\bm h$, $\bm \gamma_{L}^{\sigma}$, and $\bm \gamma_{R}$ transform as follows:
\begin{align}
&{\bm X}{\bm h}{\bm X}=\left( {\bm P}{\bm h}{\bm P} \right)^*,\label{eq:Xh}\\
&{\bm X}{\bm \gamma}_L^{\uparrow}{\bm X}={\bm \gamma}_L^{\downarrow},\label{eq:Xg}\\
&{\bm X}{\bm \gamma}_R{\bm X}={\bm \gamma}_R\label{eq:Xg2},
\end{align}
where 
\begin{align}
{\bm X}&=
\begin{pmatrix}
0&0&0&0&1&0&0&0\\
0&0&0&0&0&1&0&0\\
0&0&0&0&0&0&1&0\\
0&0&0&0&0&0&0&1\\
1&0&0&0&0&0&0&0\\
0&1&0&0&0&0&0&0\\
0&0&1&0&0&0&0&0\\
0&0&0&1&0&0&0&0
\end{pmatrix},\\
{\bm P}&=
\begin{pmatrix}
1&0&0&0&0&0&0&0\\
0&-1&0&0&0&0&0&0\\
0&0&1&0&0&0&0&0\\
0&0&0&-1&0&0&0&0\\
0&0&0&0&-1&0&0&0\\
0&0&0&0&0&1&0&0\\
0&0&0&0&0&0&-1&0\\
0&0&0&0&0&0&0&1
\end{pmatrix}.
\end{align}

From Eqs. \eqref{eq:Xh}, \eqref{eq:Xg} and \eqref{eq:Xg2}, 
transformation of ${\bm G}^{\uparrow}(\epsilon)$ under ${\bm X}$ can be written as
\begin{align}
{\bm X}{\bm G}^{r\uparrow}(\epsilon){\bm X} 
&={\bm X} \left( \epsilon-{\bm h}-i\frac{ {\bm \gamma}_L^{\uparrow}+{\bm \gamma}_R}{2} \right)^{-1} {\bm X}\nonumber \\
&=\left( \epsilon-({\bm P}{\bm h}{\bm P})^*-i\frac{ {\bm \gamma}_L^{\downarrow}+{\bm \gamma}_R}{2} \right)^{-1} \nonumber\\
&=\left ( {\bm P}\left( \epsilon-{\bm h}+i\frac{ {\bm \gamma}_L^{\downarrow}+{\bm \gamma}_R}{2} \right)^{-1} {\bm P}\right )^{*} \nonumber \\
&=\left ( {\bm P} {\bm G}^{a\downarrow}(\epsilon) {\bm P} \right )^{*} \nonumber\\
&= \left( {\bm P} {\bm G}^{r\downarrow}(\epsilon) {\bm P} \right)^{T}.\label{eq:trans}
\end{align}
Using the relations 
\begin{align}
\left [{\bm X}{\bm G}^{r\uparrow}(\epsilon){\bm X} \right]_{il{\bar \sigma},i'l'{\bar \sigma'}}
&=\left [{\bm G}^{r\uparrow}(\epsilon)\right ]_{il\sigma,i'l'\sigma'},\\
\left [ \left({\bm P}{\bm G}^{r\downarrow}(\epsilon){\bm P}\right)^T \right]_{il{\bar \sigma},i'l'{\bar \sigma'}}
&=\left [{\bm P}{\bm G}^{r\downarrow}(\epsilon){\bm P}\right ]_{i'l'{\bar \sigma'},il{\bar \sigma}}
\end{align}
together with Eq. \eqref{eq:trans}, we obtain 
\begin{eqnarray}
 \left |G_{il\sigma,i'l'\sigma'}^{\uparrow r} (\epsilon)\right |=\left |G^{\downarrow r}_{i'l'\bar{\sigma}',il\bar{\sigma} }(\epsilon)\right |.
\end{eqnarray}
Eq. \eqref{eq:Xh} (and consequently Eq.~\eqref{eq:lemma}) remains valid even in the presence of intersite SOC terms and additional hopping terms of the form $c_{2a}^\dagger c_{1b\sigma} + \mathrm{c.c.}$ and $c_{2b}^\dagger c_{1a\sigma} + \mathrm{c.c.}$. 

\begin{comment}
\section{{Proof of \eqref{eq:nn}} }
We prove $n_{1b{\downarrow},1b{\downarrow}}^\uparrow \neq n_{1b{\uparrow},1b{\uparrow}}^\downarrow$ for $U_{a}=U_{b}=U_{ab}^s=U_{ab}^t=0$, $\gamma_b=0$, $f_L=1$ and $f_R=0$.
\begin{align}
&{\bm n}^\sigma=\frac{1}{2\pi}\int_{-\infty}^{\infty} {\bm G}^r(\epsilon)({\bm \gamma}_L^\sigma f_L + {\bm \gamma}_R f_R ){\bm G}^a(\epsilon)d\epsilon
\end{align}
\end{comment}

\bibliography{ref} % Produces the bibliography via BibTeX.

\end{document}